\documentclass[twocolumn,superscriptaddress,showpacs,notitlepage]{revtex4-1}
\usepackage{amsmath,amssymb}
\usepackage{graphicx}
\usepackage{txfonts}
\usepackage{color}
\usepackage{hyperref}
\usepackage{bbold}
\usepackage{framed} 
\usepackage{wrapfig}
\usepackage[a4paper]{geometry}
\global\long\def\av#1{\left\langle #1 \right\rangle }

\global\long\def\ha{\hat{a}}
\global\long\def\had{\hat{a}^\dag}
\global\long\def\hb{\hat{b}}
\global\long\def\hbd{\hat{b}^\dag}

\global\long\def\had{\hat{a}^\dag}

\global\long\def\eps{\varepsilon}

\global\long\def\deps{\delta\varepsilon_0}

\begin{document}

\title{Critical behavior at dynamical phase transition in the generalized Bose-Anderson model}

\author{Dmitry V. Chichinadze}
\email[Current address:  School of Physics and Astronomy,
University of Minnesota, Minneapolis, MN 55455, USA,  ]{chich013@umn.edu}
\affiliation{Russian Quantum Center, Novaya 100, 143025 Skolkovo, Moscow Region, Russia}
\affiliation{Department of Physics, Lomonosov Moscow State University, Leninskie gory 1, 119991 Moscow, Russia}
\author{Alexey N. Rubtsov}
\affiliation{Russian Quantum Center, Novaya 100, 143025 Skolkovo, Moscow Region, Russia}
\affiliation{Department of Physics, Lomonosov Moscow State University, Leninskie gory 1, 119991 Moscow, Russia}

\begin{abstract}
Critical properties of the dynamical phase transition in the quenched generalized Bose-Anderson impurity model are studied in the mean-field limit of an infinite number of channels. The transition separates the evolution towards ground state and towards the  branch of stable excited states. We perform numerically exact simulations of a close vicinity of the critical quench amplitude.
The relaxation constant describing the asymptotic evolution towards ground state, as well as asymptotic 
frequency of persistent phase rotation and number of cloud particles at stable excited state are power functions of the detuning from the critical quench amplitude. The critical evolution (separatrix between the two regimes) shows a non-Lyapunov  power-law instability arising after a certain critical time. The observed critical behavior is attributed to the irreversibility of the dynamics of particles leaving the cloud and to memory effects related to the low-energy behavior of the lattice density of states.\end{abstract}

\maketitle

The dynamics of correlated quantum systems that show phase transitions at their equilibrium became a subject of intensive investigations in the last decade.  One of the key differences between the classical and quantum ensembles is that the later can emerge the undamped (persistent) excitations such as, for example, vertexes in the superfluids and superconductors \cite{Warner2005,Yuzbashyan2006,Yuzbashyan2015}. Therefore, whereas a generic nonlinear classical system at finite temperature is ergodic, a quantum fluid is not necessarily. This gives rise to phase transitions seen in the asymptotic dynamics of an open quantum system: dependent on initial conditions, it can either relax to its ground state or not.  The two regimes are separated by the dynamical transition point. In a wider context, dynamical transitions are singularities arising in the many body quantum dynamics. In particular, several models show a critical time $t^*$ after which an instability develops   \cite{Heyl2013,Heyl2015,Mitra2012}.

The studies of how the criticality shows up in the dynamics have a long history but firstly have been mostly devoted to classical systems. It was found that the dynamical critical exponents arise (see \cite{Odor2004,Sachdev2007}). The most known is the $Z$-index relating the correlation length $\xi$ and the relaxation time $\tau$ via $\tau\propto \xi^Z$ \cite{Hohenberg1977,Odor2004}. The values of dynamical indexes cannot be expressed via static ones (that is, a dynamical effective Hamiltonian obeys larger number of relevant parameters then its static counterpart).  Moreover, varying parameters of a system within the same static universality class, one can obtain different values of dynamical indexes. To describe such a situation the dynamical subclasses have been introduced \cite{Popkov2015}. 

The dynamical quantum criticality was reported in a number of recent works \cite{Heyl2013,Heyl2015,Pollmann2010} where the dynamics 
of transverse-field Ising model in low dimensions has been studied. 
Loschmidt echo rate scaling \cite{Heyl2015} was described using the renormalization group 
performed in a complex parameter space. However, universality of this procedure has been questioned in a very recent work \cite{Karrasch2017}. Unfortunately the results 
of this work are also very model-specific as the after-quench Hamiltonian is purely classical. 
A generic evidence about the criticality emerged in quantum dynamics remains very limited because of numerical issues.
Quantum Monte Carlo (QMC) method is almost the only numerically exact approach to simulate a generic system on a lattice.  However the long-time QMC calculations of high accuracy are complicated  because of the growing sign problem. Nevertheless dynamical transitions have been observed in QMC calculations of Hubbard-like models \cite{Eckstein2009,Strand2015} although the dynamical scaling was not studied.

Phase transitions are usually associated with translationally-invariant systems, but in quantum case they can also emerge in so-called impurity problems, that is, in localized (zero-dimensional) nonlinear systems coupled to a Gaussian thermostat. In particular this is the case for the Bose-Anderson model \cite{Lee2007}, which describes a   2-4 nonlinear oscillator connected to the Gaussian lattice obeying a power-law density of states. This model is closely related to the celebrated Dicke model \cite{Dicke1954,Baumann2010}. Although the nonlinearity is localized at a single spatial point, a spontaneous symmetry breaking of the ground state can occur \cite{Lee2010,Warnes2012}.  
The phase diagram contains high- and low-symmetry phases, respectively called local Mott insulator (lMI) and local Bose-Einstein condensate (lBEC).

In our recent paper \cite{Chichinadze2016}, we have shown that impurity models can exhibit dynamical phase transitions as well. 
We have studied the generalized Bose-Anderson model, in which 
$N$ identical  nonlinear oscillators are connected to the same lattice site.
For $N\to\infty$ the mean-field treatment of the model becomes exact, allowing for a simple numerical handling of 
its real-time dynamics. It was found that the symmetry broken state of the system, being subjected to a quench of parameters, either relaxes to the new ground state or reaches a stable excited state, dependent on the quench amplitude. The two asymptotic regimes are separated by a dynamical transition.

In the present paper, we perform a systematic study of the quenched  generalized Bose-Anderson model and conclude that the dynamical phase transition is a generic property of the lBEC phase. 
Further we investigate a vicinity of the dynamical transition. We detect the power-law dependence of the asymptotic evolution characteristics from the detuning of the quench amplitude  from its critical value.
Moreover, after a critical time $t^*$ the evolution itself appears to be unstable with a power-law type of instability. Our numerics suggests that the observed critical indexes are simple fractions, as one would expect for an $N\to \infty$ case.  

The Hamiltonian of the generalized Bose-Anderson model reads \cite{Chichinadze2016}
\begin{equation}
\label{hamiltonian}
\hat{H}=\sum_j H_{\mathrm{SI}} [\had_j  \ha_j] - \sum_{j, k} \frac{V}{\sqrt{N}} (\had_j \hb_k + \hbd_k \ha_j ) + \sum_k \epsilon_k \hbd_k \hb_k,
\end{equation}
where
\begin{equation}
\hat{H}_{\mathrm{SI}} [\had_j \ha_j] = \varepsilon_0 \had_j \ha_j + \frac{1}{2} \had_j \had_j  \ha_j \ha_j 
\end{equation}
is the Hamiltonian of a single component of the impurity, $\had, \ha$ and $\hbd, \hb$ are creation-annihilation operators acting at impurities numbered with $j$ and lattice modes numbered with $k$ respectively, 
$\varepsilon_0$ is the impurity on-site potential,  and $V$ determines the coupling between impurities and the lattice. Following our previous paper we consider a cubic lattice, so that $\epsilon_{k}=2h(3-\mathrm{cos}(k_x) - \mathrm{cos}(k_y) - \mathrm{cos}(k_y))$, and assume $h=1$. The ${\mathbf k}=0$ mode is excluded from Hamiltonian to remove the effects related to Bose-Einstein condensation in the bulk of the lattice. 

\begin{figure}[t]
 \center{\includegraphics[width=1\linewidth]{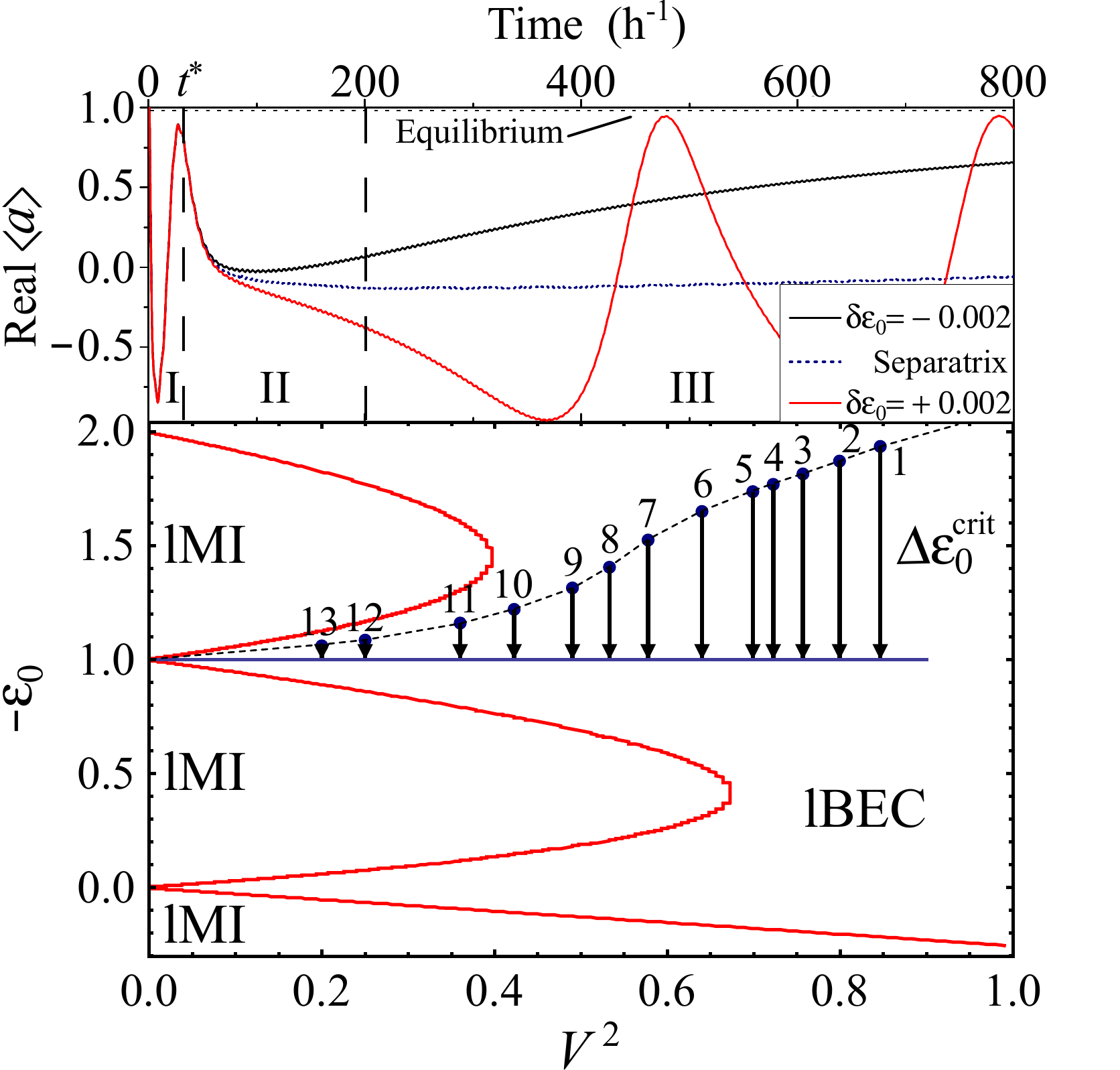}} 
\centering{}\caption{Upper panel: time evolution of real part of the order parameter for three different quench amplitudes. Dotted line depicts time evolution for critical quench amplitude (separatrix), black and red lines correspond to quenches with detuning
$\deps=\pm0.002$ from the critical value. Black line corresponds to a solution relaxing to the new equilibrium state while red 
line corresponds to a stable excited state (rotating phase).
Lower panel: critical quenches plotted on an equilibrium phase diagram of the model. Dashed line and dots depict positions on phase diagram from which critical quench 
takes place.}
\label{figvam} 
\end{figure}

In the limit of $N\to \infty$, the effect of lattice is reduced to a classical field $\lambda$ acting on the impurity,
so that the system is described by the effective Hamiltonian
\begin{equation}
\label{eff_hamiltonian}
H^{\mathrm{eff}}=\varepsilon_0 \had  \ha + \frac{1}{2} \had \had  \ha \ha - \lambda \had - \lambda^* \ha.
\end{equation}
In the symmetry broken lBEC phase the self-consistency condition 
\begin{equation}\label{equilibrium}
\lambda_{eq}= \sum_k \frac{V^2}  {\epsilon_k} \av{a}
\end{equation}
holds with a non-zero order parameter $ \av{a}$; the average is taken over the ground state of (\ref{eff_hamiltonian}) with $\lambda_{eq}$ substituted. For lMI phase this equation is trivially fulfilled with $\av{a}=0, \; \lambda_{eq}=0.$  The equilibrium phase diagram of the model is shown in the lower panel of Figure \ref{figvam}. We study a type of quenches for which a dynamical transition was reported in \cite{Chichinadze2016}. They correspond to a sudden lowering $-\varepsilon_0$ within the lBEC domain for the system initially prepared in its ground state.

Out of equilibrium the field $\lambda$ is time-dependent and has a memory about the past:
\begin{equation}
\label{lambda_t}
\lambda(t)=\sum_k V^2 \frac{e^{-i \epsilon_k t}}{\epsilon_k} \av{a(0)} + i \sum_k V^2 \int_0^t \av{a(t')} e^{-i \epsilon_k (t-t')}dt'.
\end{equation}

Possible asymptotic states of the system described by (\ref{lambda_t}) are either the new equilibrium state,
or the stable excited state with a persistently rotating phase $\av{a(t)}=a_\infty e^{i \omega t}$.
The value of $a_\infty$ can be found by switching to the rotation frame \cite{Chichinadze2016}, where the chemical potential appears to be shifted; $\varepsilon_0 \to \varepsilon_0+ \omega$, $\epsilon_k \to \epsilon_k+ \omega$. After this shift is accounted, $a_\infty$ satisfies the ``equilibrium'' equations (\ref{eff_hamiltonian}, \ref{equilibrium}). As it is immediately seen,  $\omega$ must be positive: the chemical potential shift causes instability otherwise.  
An appearance of the stable excited states is related to the relaxation mechanism of the model:
the excited system loses its energy by the emission of bosons away from 
the impurity and the surrounding cloud \cite{Chichinadze2016}.  Stable excited states have less particles then the equilibrium state and therefore cannot relax. 
In particular, an increase of $-\eps_0$ always results in an excited state, as the new ground state requires a larger amount of particles than there exists in the system.

Lowering $-\eps_0$, one observes the quenched dynamics towards one of the two different asymptotic regimes.
For a small positive quench amplitude $\Delta\varepsilon_{0}=\varepsilon_{0}(t>0)-\varepsilon_{0}(t<0)$, loss of particles results in evolution towards the new ground state.
 Increasing $\Delta\varepsilon_0$, with no change in $\eps_0(t>0)$, results in the formation of the persistently rotating phase.
The two scenarios are separated by the dynamical transition singularity occurred at certain $\Delta\varepsilon_{0}^{\mathrm{crit}}$.

The lower panel of Figure \ref{figvam} shows critical quenches plotted on the equilibrium phase diagram. We have performed simulations for a number of final quench points $(V, -\eps_0)$ along the line $-\eps_0=1$ separating two local Mott lobes. The described dynamical transition was observed for all points including whose lying between the lobes: smaller $V$ corresponds to narrower lBEC strip, and also to smaller critical quench value.  Inside the lBEC phase, the transition does not show a link to the equilibrium phase diagram and are therefore interpreted as purely dynamical phenomenon. This makes quenches within lBEC different from whose involving lMI phase \cite{Chichinadze2016}, whose transient dynamics is closely related with the equibrium phases on the quench path.

\begin{figure}[t]
\center{\includegraphics[width=1\linewidth]{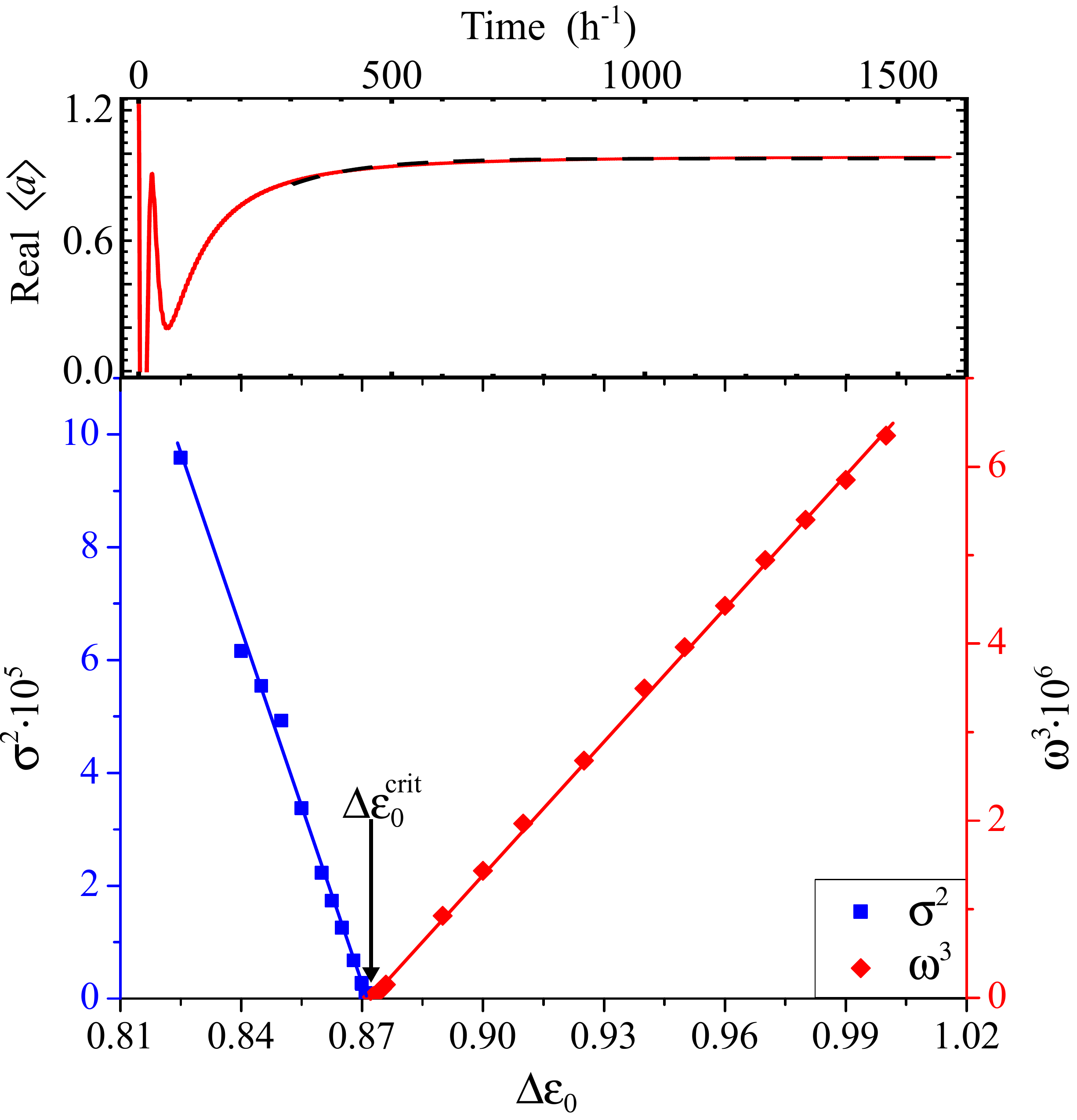}} 
\centering{}\caption{Upper panel: exponential fit of equilibration asymptotic dynamics. Red curve shows the dynamics of real part of the order 
parameter while black dashed one depicts exponential fit. Lower panel: power-law dependences of relaxation parameter $\sigma$ (blue dots) and frequency $\omega$ (red dots) as functions of quench amplitude $\Delta\varepsilon_0$. In this picture critical quench amplitude
is $\Delta\varepsilon_0^{\mathrm{crit}}=0.8719$. As it is seen from the picture, $\sigma \propto (\deps)^{1/2}$ and
$\omega \propto (\deps)^{1/3}$.
}
\label{figsdva} 
\end{figure}

In our study we examined a close vicinity of the critical quenches. Let us introduce the deviation of the quench amplitude from the critical value $\deps=\Delta\varepsilon_{0}^{\mathrm{crit}}-\Delta\eps_0$; negative (positive) $\deps$ corresponds to the quenches below (above) the critical value.  In our calculations, values of critical quench amplitudes were estimated with an accuracy of at least $10^{-3}$.
The upper panel of Figure \ref{figvam}  shows time dependence of the order parameter for three quench amplitudes: right at the dynamical transition ($\deps=0$ -- this curve can be called separatrix)), slightly above and slightly below the critical quench value. A closer look at the numerical data shows that the three characteristic time intervals can be introduced.
Within the interval I, up to critical time $t^*$ (indicated by the vertical dashed line in Figure \ref{figvam}, upper panel) the deviation $\delta a(t)=\av{a(t)}-\av{a_{\mathrm{crit}}(t)}$ does not significantly change.   For $t>t^*$, the evolution becomes unstable: $\delta a(t)$ increases with time. We introduce time interval II, where $\delta a$ grows in time, but still remains small compared to $\av{a}$. 
The asymptotic evolution is realized in the interval III, where $\delta a$ and $\av{a}$ are of the same order.
Note that whereas $t^*$ is independent of $\deps$, the crossover between the time intervals II and III occurs at larger time arguments for smaller $\deps$.

We address the following questions: (i) concerning the time interval III, how the equilibration occurs  for small negative $\deps $ and how  asymptotic rotation frequency behaves  for small positive $\deps$  and (ii) concerning the time interval II, what is the type of instability for the separatrix.

\begin{figure}[t]
\center{\includegraphics[width=1\linewidth]{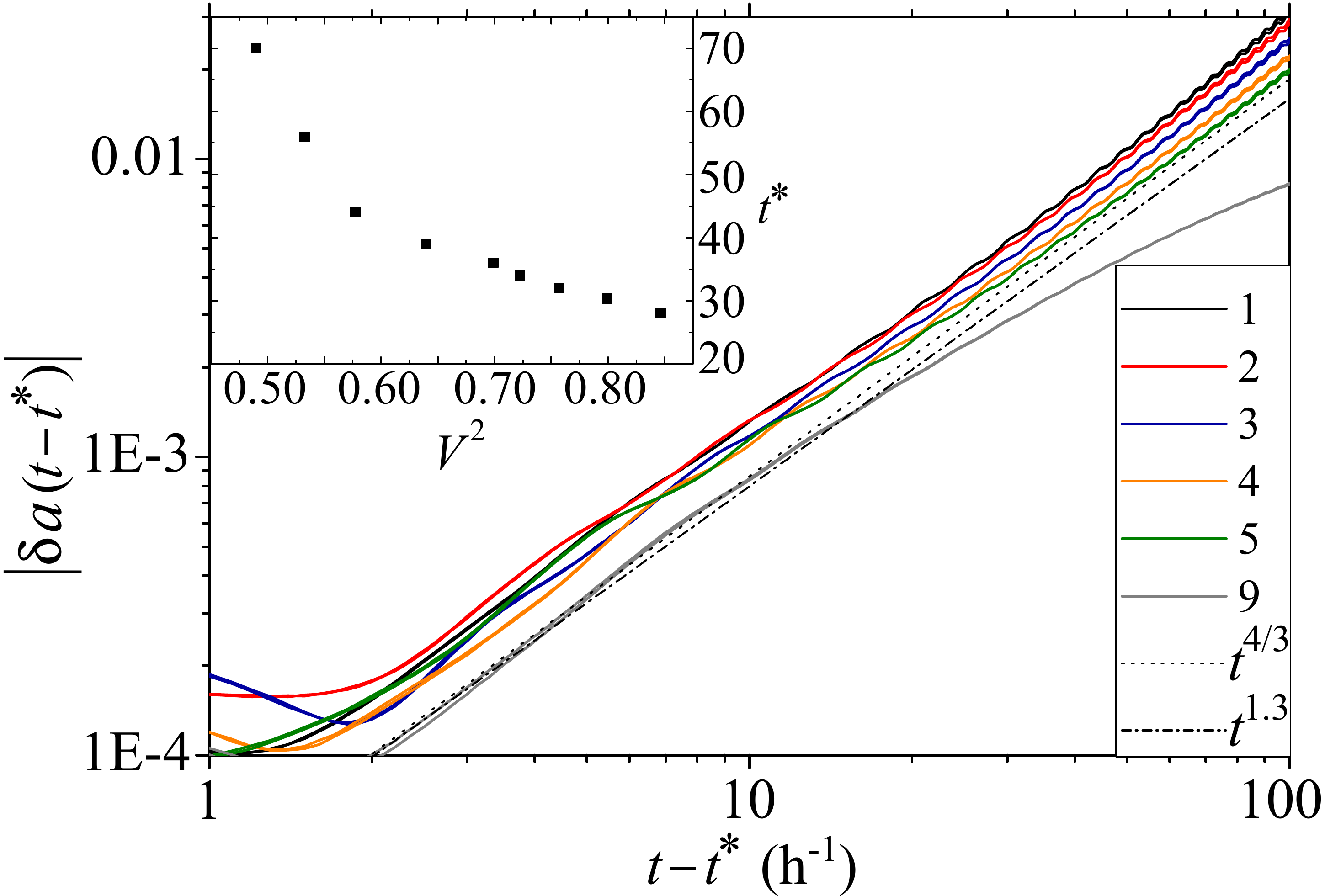}} 
\centering{}\caption{The dependence of $|\delta a(t-t^*)|=|\av{a(t-t^*)}-\av{a_{\mathrm{crit}}(t-t^*)}|$ on $t-t^*$,
where $|\av{a}|_{\mathrm{crit}}$ corresponds to critical quench with $\deps=0$.
One can consider this quantity as a divergence of trajectories in phase space (as it is commonly considered in nonlinear dynamics),
that shows the kind of instability of a critical point. Numbers of curves correspond to numbers of critical quenches shown in Fig. 1. 
Two lines with the same color (almost on top of each other) depict curves for quenches with $\deps=\pm 0.0005$. Dashed line shows a power-law fit with power $4/3$ and dash-dotted line depicts a power-law fit with power $1.3$
for these dependences. Inset: dependence of critical time $t^*$ on $V$.
}
\label{instability} 
\end{figure}

An analysis of the equilibration at $\deps<0$ have shown that the asymptotic deviation of the order parameter from its equilibrium value falls  exponentially with time. The upper panel of Figure \ref{figsdva} provides an example of the exponential fit $\av{a}(t)=a_{eq}- a_1 e^{-\sigma t}$ at  $\deps=-0.0219$ for the quench to $(V=0.894, -\eps_0=1)$  -- the quench 2 from Figure \ref{figvam}. Values of the relaxation constant $\sigma$ obtained from the fitting procedure show a square-root dependence on $|\deps|$, as the lower panel of Figure \ref{figsdva} shows:  we plot $\sigma^2$ vs. $\deps$ and observe a linear dependence. The same panel presents our results for the asymptotic rotation frequency for the system being quenched with  $\deps>0$ to the same point $(V=0.894, -\eps_0=1)$. In this case $\omega^3$ (the right axis) exhibits linear dependence, so that we conclude about $\omega\propto(\deps)^{1/3}$.

Our conclusions allow to estimate how the number of particles asymptotically remaining in the lBEC cloud $N_{lBEC}$ scales with $\deps$. The rotation-frame analysis \cite{Chichinadze2016} gives the expression $N_{lBEC}\propto \int \frac{d^3 k}{(\omega+\epsilon_k)^2}$.  It  leads to the divergence $N_{lBEC}\propto \omega^{-1/2}$. 
We consequently conclude about the scaling law $N_{lBEC}\propto (\deps)^{-1/6}$ at small positive  $\deps$.

Now let us turn to the analysis of the instability of the critical evolution at $t>t^*$, i.e. in the interval II. 
Figure \ref{instability} presents a log-log plot of $|\delta a(t-t^*)|$ calculated for a set of quenches with different impurity-lattice coupling $V$; the values of $t^*$ vs. $V$ are plotted in the inset.
For each quench, we have considered two values of $\deps=\pm 5\cdot10^{-4}$. It appears that the sign of $\deps$ almost does not affect the absolute value of  $|\delta a|$. For each quench Figure \ref{instability} shows two dependencies of $|\delta a(t-t^*)|$, that correspond to the positive and negative $\deps$, but these two dependencies are almost indistinguishable from each other. Besides small high-frequency oscillations,   
possibly related to the upper cut-off introduced by a discrete lattice,  the log-log curves shown in Figure \ref{instability} remain linear while the time argument is changed by 2 orders of magnitude; however this range is narrowed for the quench 9 performed for the smallest $V$, i.e. closer to the lMI region.

We conclude about the power-law dynamical instability $\ln |\delta a| \propto \ln (t-t^*)$ with the 
index being the same for all quenches. We remind that the usual scenario known from the nonlinear dynamics of classical finite systems is the Lyapunov instability  \cite{Strogatz2014}: a small deviation from the separatrix trajectory grows in time exponentially. For the system considered, this does not appear to be the case. 

The exact index value extracted from the numerical data slightly depends on how the values of $t^*$ are chosen. A finite value of $\deps$ used in calculations corresponds to a finite region where a stable evolution passes into the power-law instability; this results in certain errorbar for the value $t^*$. We have performed estimations of the index for $t^*$ taken within this region and obtained the index close to 1.30 with an errorbar of about 0.03.   The mean-value 1.30, as well as the log-log dependencies shown in Figure \ref{instability}, are obtained for $t^*$ fitted to minimize the crossover region between the stable and power-law evolution.
 For comparison, we have drawn straight lines in Figure \ref{instability} corresponding to the indexes 1.3 and 4/3.

We interpret the index obtained as a critical exponent of the dynamical transition occurring at $t^*$.  Additional comments are needed at this point, because it is not a priori clear if the concept of criticality is applicable for the system studied. Indeed, transitions at equilibrium, for which the theory of criticality was initially developed, occur in (in average) uniform systems, whereas we deal  with a localized one.  Another seminal example of a critical behavior is known from the Feigenbaum's \cite{Feigenbaum1979, Strogatz2014} theory for the transition between the periodic and chaotic one-dimensional mapping. In this case, the localized in space system is out of equilibrium, but the very difference from our case is that the Feigenbaum's theory describes a steady process. Nevertheless our data reported above suggest that our system shows a critical behavior. 
The power-law instability signals that the phenomenon exhibits collective behavior: a few-body dynamics at finite time scale would be likely characterized by a Lyapunov exponent \cite{Strogatz2014}. In this regard we emphasize that the memory kernel of the lattice response in the Equation (\ref{lambda_t}) obeys a long memory which makes the mathematical difference 
of our model from a finite set of differential equations common for nonlinear dynamics.
Furthermore, the very same dependence of  $|\delta a|$ on $|\delta \varepsilon_0|$ on different sides of the transition reported in
Figure \ref{instability}, excludes scenarios related to locking of some states etc. On the other hand, it resembles the transitions at equilibrium, when   fluctuations show the same behavior below and above the transition point.

Other property  worth to discuss is that the dynamical transition studied is not followed by a symmetry change: the order parameter does not equal zero neither at the lBEC ground state, nor in the state with the persistently rotating phase. The latter can be seen as a ground state of the
Hamiltonian in the rotating frame, where the spectrum is gapped with the gap $\omega$. The situation when 
the two states have the same symmetry, but one of them is gapped and another is not, is known for the equilibrium quantum phase transitions. For example, this is the case for Mott metal-insulator transitions \cite{Mott1990,Furukawa2015} (here we  refer to the Mott transition which does not break the translational symmetry and leave out possible antiferromagnetism in Mott systems). As well, equilibrium quantum transitions occurring inside a symmetry broken phase are known: an example is ferroelectic transitions in the crystals lacking inversion symmetry (e.g. in $\mathrm{ArCrS_{2}}$ \cite{Streltsov2015}).
 
Finally, possible experimental relevance is to be discussed. In this context we consider of high importance the finding that the dynamical transition was observed in a large part of the lBEC phase, including the weak coupling limit $V\to 0$.
In this limit generalized Bose-Anderson Hamiltonian is closely related to Dicke model \cite{Dicke1954,Yuzbashyan2015}, as just two impurity levels are important. That is, the dynamical transition can be looked for in qubit systems where spontaneous symmetry breaking emerges (provided proper density of photon states). Another possibility is to operate with ultracold atoms. In this case, a Josephson current \cite{Levy2007}  between the lBECs of two systems with slightly different quench values can be used to measure their relative phase.

Let us present some preliminary argumentation about the applicability of our results to finite-$N$ systems, although a consideration beyond mean-field is out of the scope of this paper.
We expect that the dynamical transition studied can also be found away from mean-field, in the impurity systems obeying two properties: (i) the symmetry broken phase is characterized by an infinite number of particles in the lBEC cloud surrounding the impurity and (ii) a irreversibility of dynamics of the particles  leaving the cloud, which gives rise to stable excited states having less particles and higher energy then the ground state. The first property was indeed observed in the NRG calculations for the single-impurity Bose-Anderson model \cite{Lee2010}. The second is a simple property generic to all systems where the particles evaporate in vacuum. Quantitative properties of the transition, such as index values, can be altered from the  $N\to\infty$.
However we mention that in some cases even index values found from mean-field grounds appear to be quite accurate; a good example is the Flory description of self-avoiding chains \cite{Flory1969}.

To summarize, using the mean-field approach we have studied the dynamical phase transition, arising within the symmetry broken phase of the generalized Bose-Anderson impurity model. We have  studied a vicinity of the transition point and found that characteristics of the asymptotic evolution (namely, the relaxation parameter and the frequency of persistent oscillations) are power functions of the detuning from the critical quench amplitude. Furthermore the critical evolution (separatrix) also shows a non-Lyapunov  power-law instability, arising after a critical time $t^*$. We attribute the observed phenomena to the irreversibility of the dynamics of particles leaving the lBEC cloud and to the memory effects related to the low-energy behavior of the lattice density of states.

Acknowledgments. We acknowledge useful discussions with Pedro Ribeiro, Georg Rohringer and Yulia Shchadilova. Authors thank the Dynasty foundation
and RFBR Grant No. 16-32-00554 for financial support.

\bibliography{biblio}

\begin{thebibliography}{27}%
\makeatletter
\providecommand \@ifxundefined [1]{%
 \@ifx{#1\undefined}
}%
\providecommand \@ifnum [1]{%
 \ifnum #1\expandafter \@firstoftwo
 \else \expandafter \@secondoftwo
 \fi
}%
\providecommand \@ifx [1]{%
 \ifx #1\expandafter \@firstoftwo
 \else \expandafter \@secondoftwo
 \fi
}%
\providecommand \natexlab [1]{#1}%
\providecommand \enquote  [1]{``#1''}%
\providecommand \bibnamefont  [1]{#1}%
\providecommand \bibfnamefont [1]{#1}%
\providecommand \citenamefont [1]{#1}%
\providecommand \href@noop [0]{\@secondoftwo}%
\providecommand \href [0]{\begingroup \@sanitize@url \@href}%
\providecommand \@href[1]{\@@startlink{#1}\@@href}%
\providecommand \@@href[1]{\endgroup#1\@@endlink}%
\providecommand \@sanitize@url [0]{\catcode `\\12\catcode `\$12\catcode
  `\&12\catcode `\#12\catcode `\^12\catcode `\_12\catcode `\%12\relax}%
\providecommand \@@startlink[1]{}%
\providecommand \@@endlink[0]{}%
\providecommand \url  [0]{\begingroup\@sanitize@url \@url }%
\providecommand \@url [1]{\endgroup\@href {#1}{\urlprefix }}%
\providecommand \urlprefix  [0]{URL }%
\providecommand \Eprint [0]{\href }%
\providecommand \doibase [0]{http://dx.doi.org/}%
\providecommand \selectlanguage [0]{\@gobble}%
\providecommand \bibinfo  [0]{\@secondoftwo}%
\providecommand \bibfield  [0]{\@secondoftwo}%
\providecommand \translation [1]{[#1]}%
\providecommand \BibitemOpen [0]{}%
\providecommand \bibitemStop [0]{}%
\providecommand \bibitemNoStop [0]{.\EOS\space}%
\providecommand \EOS [0]{\spacefactor3000\relax}%
\providecommand \BibitemShut  [1]{\csname bibitem#1\endcsname}%
\let\auto@bib@innerbib\@empty
\bibitem [{\citenamefont {Warner}\ and\ \citenamefont
  {Leggett}(2005)}]{Warner2005}%
  \BibitemOpen
  \bibfield  {author} {\bibinfo {author} {\bibfnamefont {G.~L.}\ \bibnamefont
  {Warner}}\ and\ \bibinfo {author} {\bibfnamefont {A.~J.}\ \bibnamefont
  {Leggett}},\ }\href {\doibase 10.1103/PhysRevB.71.134514} {\bibfield
  {journal} {\bibinfo  {journal} {Physical Review B}\ }\textbf {\bibinfo
  {volume} {71}},\ \bibinfo {pages} {134514} (\bibinfo {year}
  {2005})}\BibitemShut {NoStop}%
\bibitem [{\citenamefont {Yuzbashyan}\ \emph {et~al.}(2006)\citenamefont
  {Yuzbashyan}, \citenamefont {Tsyplyatyev},\ and\ \citenamefont
  {Altshuler}}]{Yuzbashyan2006}%
  \BibitemOpen
  \bibfield  {author} {\bibinfo {author} {\bibfnamefont {E.~A.}\ \bibnamefont
  {Yuzbashyan}}, \bibinfo {author} {\bibfnamefont {O.}~\bibnamefont
  {Tsyplyatyev}}, \ and\ \bibinfo {author} {\bibfnamefont {B.~L.}\ \bibnamefont
  {Altshuler}},\ }\href {\doibase 10.1103/PhysRevLett.96.097005} {\bibfield
  {journal} {\bibinfo  {journal} {Physical Review Letters}\ }\textbf {\bibinfo
  {volume} {96}},\ \bibinfo {pages} {097005} (\bibinfo {year}
  {2006})}\BibitemShut {NoStop}%
\bibitem [{\citenamefont {Yuzbashyan}\ \emph {et~al.}(2015)\citenamefont
  {Yuzbashyan}, \citenamefont {Dzero}, \citenamefont {Gurarie},\ and\
  \citenamefont {Foster}}]{Yuzbashyan2015}%
  \BibitemOpen
  \bibfield  {author} {\bibinfo {author} {\bibfnamefont {E.~A.}\ \bibnamefont
  {Yuzbashyan}}, \bibinfo {author} {\bibfnamefont {M.}~\bibnamefont {Dzero}},
  \bibinfo {author} {\bibfnamefont {V.}~\bibnamefont {Gurarie}}, \ and\
  \bibinfo {author} {\bibfnamefont {M.~S.}\ \bibnamefont {Foster}},\ }\href
  {\doibase 10.1103/PhysRevA.91.033628} {\bibfield  {journal} {\bibinfo
  {journal} {Physical Review A}\ }\textbf {\bibinfo {volume} {91}},\ \bibinfo
  {pages} {033628} (\bibinfo {year} {2015})}\BibitemShut {NoStop}%
\bibitem [{\citenamefont {Heyl}\ \emph {et~al.}(2013)\citenamefont {Heyl},
  \citenamefont {Polkovnikov},\ and\ \citenamefont {Kehrein}}]{Heyl2013}%
  \BibitemOpen
  \bibfield  {author} {\bibinfo {author} {\bibfnamefont {M.}~\bibnamefont
  {Heyl}}, \bibinfo {author} {\bibfnamefont {A.}~\bibnamefont {Polkovnikov}}, \
  and\ \bibinfo {author} {\bibfnamefont {S.}~\bibnamefont {Kehrein}},\ }\href
  {\doibase 10.1103/PhysRevLett.110.135704} {\bibfield  {journal} {\bibinfo
  {journal} {Phys. Rev. Lett.}\ }\textbf {\bibinfo {volume} {110}},\ \bibinfo
  {pages} {135704} (\bibinfo {year} {2013})}\BibitemShut {NoStop}%
\bibitem [{\citenamefont {Heyl}(2015)}]{Heyl2015}%
  \BibitemOpen
  \bibfield  {author} {\bibinfo {author} {\bibfnamefont {M.}~\bibnamefont
  {Heyl}},\ }\href {\doibase 10.1103/PhysRevLett.115.140602} {\bibfield
  {journal} {\bibinfo  {journal} {Phys. Rev. Lett.}\ }\textbf {\bibinfo
  {volume} {115}},\ \bibinfo {pages} {140602} (\bibinfo {year}
  {2015})}\BibitemShut {NoStop}%
\bibitem [{\citenamefont {Mitra}(2012)}]{Mitra2012}%
  \BibitemOpen
  \bibfield  {author} {\bibinfo {author} {\bibfnamefont {A.}~\bibnamefont
  {Mitra}},\ }\href {\doibase 10.1103/PhysRevLett.109.260601} {\bibfield
  {journal} {\bibinfo  {journal} {Phys. Rev. Lett.}\ }\textbf {\bibinfo
  {volume} {109}},\ \bibinfo {pages} {260601} (\bibinfo {year}
  {2012})}\BibitemShut {NoStop}%
\bibitem [{\citenamefont {\'Odor}(2004)}]{Odor2004}%
  \BibitemOpen
  \bibfield  {author} {\bibinfo {author} {\bibfnamefont {G.}~\bibnamefont
  {\'Odor}},\ }\href {\doibase 10.1103/RevModPhys.76.663} {\bibfield  {journal}
  {\bibinfo  {journal} {Rev. Mod. Phys.}\ }\textbf {\bibinfo {volume} {76}},\
  \bibinfo {pages} {663} (\bibinfo {year} {2004})}\BibitemShut {NoStop}%
\bibitem [{\citenamefont {Sachdev}(2007)}]{Sachdev2007}%
  \BibitemOpen
  \bibfield  {author} {\bibinfo {author} {\bibfnamefont {S.}~\bibnamefont
  {Sachdev}},\ }\href@noop {} {\emph {\bibinfo {title} {Quantum phase
  transitions}}}\ (\bibinfo  {publisher} {Wiley Online Library},\ \bibinfo
  {year} {2007})\BibitemShut {NoStop}%
\bibitem [{\citenamefont {Hohenberg}\ and\ \citenamefont
  {Halperin}(1977)}]{Hohenberg1977}%
  \BibitemOpen
  \bibfield  {author} {\bibinfo {author} {\bibfnamefont {P.~C.}\ \bibnamefont
  {Hohenberg}}\ and\ \bibinfo {author} {\bibfnamefont {B.~I.}\ \bibnamefont
  {Halperin}},\ }\href {\doibase 10.1103/RevModPhys.49.435} {\bibfield
  {journal} {\bibinfo  {journal} {Rev. Mod. Phys.}\ }\textbf {\bibinfo {volume}
  {49}},\ \bibinfo {pages} {435} (\bibinfo {year} {1977})}\BibitemShut
  {NoStop}%
\bibitem [{\citenamefont {Popkov}\ \emph {et~al.}(2015)\citenamefont {Popkov},
  \citenamefont {Schadschneider}, \citenamefont {Schmidt},\ and\ \citenamefont
  {Sch\"{u}tz}}]{Popkov2015}%
  \BibitemOpen
  \bibfield  {author} {\bibinfo {author} {\bibfnamefont {V.}~\bibnamefont
  {Popkov}}, \bibinfo {author} {\bibfnamefont {A.}~\bibnamefont
  {Schadschneider}}, \bibinfo {author} {\bibfnamefont {J.}~\bibnamefont
  {Schmidt}}, \ and\ \bibinfo {author} {\bibfnamefont {G.~M.}\ \bibnamefont
  {Sch\"{u}tz}},\ }\href {\doibase 10.1073/pnas.1512261112} {\bibfield
  {journal} {\bibinfo  {journal} {Proceedings of the National Academy of
  Sciences}\ }\textbf {\bibinfo {volume} {112}},\ \bibinfo {pages} {12645}
  (\bibinfo {year} {2015})},\ \Eprint
  {http://arxiv.org/abs/http://www.pnas.org/content/112/41/12645.full.pdf}
  {http://www.pnas.org/content/112/41/12645.full.pdf} \BibitemShut {NoStop}%
\bibitem [{\citenamefont {Pollmann}\ \emph {et~al.}(2010)\citenamefont
  {Pollmann}, \citenamefont {Mukerjee}, \citenamefont {Green},\ and\
  \citenamefont {Moore}}]{Pollmann2010}%
  \BibitemOpen
  \bibfield  {author} {\bibinfo {author} {\bibfnamefont {F.}~\bibnamefont
  {Pollmann}}, \bibinfo {author} {\bibfnamefont {S.}~\bibnamefont {Mukerjee}},
  \bibinfo {author} {\bibfnamefont {A.~G.}\ \bibnamefont {Green}}, \ and\
  \bibinfo {author} {\bibfnamefont {J.~E.}\ \bibnamefont {Moore}},\ }\href
  {\doibase 10.1103/PhysRevE.81.020101} {\bibfield  {journal} {\bibinfo
  {journal} {Phys. Rev. E}\ }\textbf {\bibinfo {volume} {81}},\ \bibinfo
  {pages} {020101} (\bibinfo {year} {2010})}\BibitemShut {NoStop}%
\bibitem [{\citenamefont {Karrasch}\ and\ \citenamefont
  {Schuricht}(2017)}]{Karrasch2017}%
  \BibitemOpen
  \bibfield  {author} {\bibinfo {author} {\bibfnamefont {C.}~\bibnamefont
  {Karrasch}}\ and\ \bibinfo {author} {\bibfnamefont {D.}~\bibnamefont
  {Schuricht}},\ }\href@noop {} {\bibfield  {journal} {\bibinfo  {journal}
  {arXiv preprint arXiv:1701.04214}\ } (\bibinfo {year} {2017})}\BibitemShut
  {NoStop}%
\bibitem [{\citenamefont {Eckstein}\ \emph {et~al.}(2009)\citenamefont
  {Eckstein}, \citenamefont {Kollar},\ and\ \citenamefont
  {Werner}}]{Eckstein2009}%
  \BibitemOpen
  \bibfield  {author} {\bibinfo {author} {\bibfnamefont {M.}~\bibnamefont
  {Eckstein}}, \bibinfo {author} {\bibfnamefont {M.}~\bibnamefont {Kollar}}, \
  and\ \bibinfo {author} {\bibfnamefont {P.}~\bibnamefont {Werner}},\ }\href
  {\doibase 10.1103/PhysRevLett.103.056403} {\bibfield  {journal} {\bibinfo
  {journal} {Phys. Rev. Lett.}\ }\textbf {\bibinfo {volume} {103}},\ \bibinfo
  {pages} {056403} (\bibinfo {year} {2009})}\BibitemShut {NoStop}%
\bibitem [{\citenamefont {Strand}\ \emph {et~al.}(2015)\citenamefont {Strand},
  \citenamefont {Eckstein},\ and\ \citenamefont {Werner}}]{Strand2015}%
  \BibitemOpen
  \bibfield  {author} {\bibinfo {author} {\bibfnamefont {H.~U.~R.}\
  \bibnamefont {Strand}}, \bibinfo {author} {\bibfnamefont {M.}~\bibnamefont
  {Eckstein}}, \ and\ \bibinfo {author} {\bibfnamefont {P.}~\bibnamefont
  {Werner}},\ }\href {\doibase 10.1103/PhysRevX.5.011038} {\bibfield  {journal}
  {\bibinfo  {journal} {Phys. Rev. X}\ }\textbf {\bibinfo {volume} {5}},\
  \bibinfo {pages} {011038} (\bibinfo {year} {2015})}\BibitemShut {NoStop}%
\bibitem [{\citenamefont {Lee}\ and\ \citenamefont {Bulla}(2007)}]{Lee2007}%
  \BibitemOpen
  \bibfield  {author} {\bibinfo {author} {\bibfnamefont {H.-J.}\ \bibnamefont
  {Lee}}\ and\ \bibinfo {author} {\bibfnamefont {R.}~\bibnamefont {Bulla}},\
  }\href {\doibase 10.1140/epjb/e2007-00118-3} {\bibfield  {journal} {\bibinfo
  {journal} {The European Physical Journal B}\ }\textbf {\bibinfo {volume}
  {56}},\ \bibinfo {pages} {199} (\bibinfo {year} {2007})}\BibitemShut
  {NoStop}%
\bibitem [{\citenamefont {Dicke}(1954)}]{Dicke1954}%
  \BibitemOpen
  \bibfield  {author} {\bibinfo {author} {\bibfnamefont {R.~H.}\ \bibnamefont
  {Dicke}},\ }\href@noop {} {\bibfield  {journal} {\bibinfo  {journal}
  {Physical Review}\ }\textbf {\bibinfo {volume} {93}},\ \bibinfo {pages} {99}
  (\bibinfo {year} {1954})}\BibitemShut {NoStop}%
\bibitem [{\citenamefont {Baumann}\ \emph {et~al.}(2010)\citenamefont
  {Baumann}, \citenamefont {Guerlin}, \citenamefont {Brennecke},\ and\
  \citenamefont {Esslinger}}]{Baumann2010}%
  \BibitemOpen
  \bibfield  {author} {\bibinfo {author} {\bibfnamefont {K.}~\bibnamefont
  {Baumann}}, \bibinfo {author} {\bibfnamefont {C.}~\bibnamefont {Guerlin}},
  \bibinfo {author} {\bibfnamefont {F.}~\bibnamefont {Brennecke}}, \ and\
  \bibinfo {author} {\bibfnamefont {T.}~\bibnamefont {Esslinger}},\ }\href@noop
  {} {\bibfield  {journal} {\bibinfo  {journal} {Nature}\ }\textbf {\bibinfo
  {volume} {464}},\ \bibinfo {pages} {1301} (\bibinfo {year}
  {2010})}\BibitemShut {NoStop}%
\bibitem [{\citenamefont {Lee}\ \emph {et~al.}(2010)\citenamefont {Lee},
  \citenamefont {Byczuk},\ and\ \citenamefont {Bulla}}]{Lee2010}%
  \BibitemOpen
  \bibfield  {author} {\bibinfo {author} {\bibfnamefont {H.-J.}\ \bibnamefont
  {Lee}}, \bibinfo {author} {\bibfnamefont {K.}~\bibnamefont {Byczuk}}, \ and\
  \bibinfo {author} {\bibfnamefont {R.}~\bibnamefont {Bulla}},\ }\href
  {\doibase 10.1103/PhysRevB.82.054516} {\bibfield  {journal} {\bibinfo
  {journal} {Phys. Rev. B}\ }\textbf {\bibinfo {volume} {82}},\ \bibinfo
  {pages} {054516} (\bibinfo {year} {2010})}\BibitemShut {NoStop}%
\bibitem [{\citenamefont {Warnes}\ and\ \citenamefont
  {Miranda}(2012)}]{Warnes2012}%
  \BibitemOpen
  \bibfield  {author} {\bibinfo {author} {\bibfnamefont {J.}~\bibnamefont
  {Warnes}}\ and\ \bibinfo {author} {\bibfnamefont {E.}~\bibnamefont
  {Miranda}},\ }\href {\doibase 10.1140/epjb/e2012-30191-2} {\bibfield
  {journal} {\bibinfo  {journal} {The European Physical Journal B}\ }\textbf
  {\bibinfo {volume} {85}},\ \bibinfo {eid} {341} (\bibinfo {year} {2012}),\
  10.1140/epjb/e2012-30191-2}\BibitemShut {NoStop}%
\bibitem [{\citenamefont {Chichinadze}\ \emph {et~al.}(2016)\citenamefont
  {Chichinadze}, \citenamefont {Ribeiro}, \citenamefont {Shchadilova},\ and\
  \citenamefont {Rubtsov}}]{Chichinadze2016}%
  \BibitemOpen
  \bibfield  {author} {\bibinfo {author} {\bibfnamefont {D.~V.}\ \bibnamefont
  {Chichinadze}}, \bibinfo {author} {\bibfnamefont {P.}~\bibnamefont
  {Ribeiro}}, \bibinfo {author} {\bibfnamefont {Y.~E.}\ \bibnamefont
  {Shchadilova}}, \ and\ \bibinfo {author} {\bibfnamefont {A.~N.}\ \bibnamefont
  {Rubtsov}},\ }\href {\doibase 10.1103/PhysRevB.94.054301} {\bibfield
  {journal} {\bibinfo  {journal} {Phys. Rev. B}\ }\textbf {\bibinfo {volume}
  {94}},\ \bibinfo {pages} {054301} (\bibinfo {year} {2016})}\BibitemShut
  {NoStop}%
\bibitem [{\citenamefont {Strogatz}(2014)}]{Strogatz2014}%
  \BibitemOpen
  \bibfield  {author} {\bibinfo {author} {\bibfnamefont {S.~H.}\ \bibnamefont
  {Strogatz}},\ }\href@noop {} {\emph {\bibinfo {title} {Nonlinear dynamics and
  chaos: with applications to physics, biology, chemistry, and engineering}}}\
  (\bibinfo  {publisher} {Westview press},\ \bibinfo {year} {2014})\BibitemShut
  {NoStop}%
\bibitem [{\citenamefont {Feigenbaum}(1979)}]{Feigenbaum1979}%
  \BibitemOpen
  \bibfield  {author} {\bibinfo {author} {\bibfnamefont {M.~J.}\ \bibnamefont
  {Feigenbaum}},\ }\href {\doibase 10.1007/BF01107909} {\bibfield  {journal}
  {\bibinfo  {journal} {Journal of Statistical Physics}\ }\textbf {\bibinfo
  {volume} {21}},\ \bibinfo {pages} {669} (\bibinfo {year} {1979})}\BibitemShut
  {NoStop}%
\bibitem [{\citenamefont {Mott}(1990)}]{Mott1990}%
  \BibitemOpen
  \bibfield  {author} {\bibinfo {author} {\bibfnamefont {N.~F.}\ \bibnamefont
  {Mott}},\ }\href@noop {} {\emph {\bibinfo {title} {Metal-Insulator
  Transitions}}}\ (\bibinfo  {publisher} {Taylor and Francis},\ \bibinfo {year}
  {1990})\BibitemShut {NoStop}%
\bibitem [{\citenamefont {Furukawa}\ \emph {et~al.}(2015)\citenamefont
  {Furukawa}, \citenamefont {Miyagawa}, \citenamefont {Taniguchi},
  \citenamefont {Kato},\ and\ \citenamefont {Kanoda}}]{Furukawa2015}%
  \BibitemOpen
  \bibfield  {author} {\bibinfo {author} {\bibfnamefont {T.}~\bibnamefont
  {Furukawa}}, \bibinfo {author} {\bibfnamefont {K.}~\bibnamefont {Miyagawa}},
  \bibinfo {author} {\bibfnamefont {H.}~\bibnamefont {Taniguchi}}, \bibinfo
  {author} {\bibfnamefont {R.}~\bibnamefont {Kato}}, \ and\ \bibinfo {author}
  {\bibfnamefont {K.}~\bibnamefont {Kanoda}},\ }\href {\doibase
  10.1038/nphys3235} {\bibfield  {journal} {\bibinfo  {journal} {Nature
  Physics}\ }\textbf {\bibinfo {volume} {11}},\ \bibinfo {pages} {221}
  (\bibinfo {year} {2015})}\BibitemShut {NoStop}%
\bibitem [{\citenamefont {Streltsov}\ \emph {et~al.}(2015)\citenamefont
  {Streltsov}, \citenamefont {Poteryaev},\ and\ \citenamefont
  {Rubtsov}}]{Streltsov2015}%
  \BibitemOpen
  \bibfield  {author} {\bibinfo {author} {\bibfnamefont {S.~V.}\ \bibnamefont
  {Streltsov}}, \bibinfo {author} {\bibfnamefont {A.~I.}\ \bibnamefont
  {Poteryaev}}, \ and\ \bibinfo {author} {\bibfnamefont {A.~N.}\ \bibnamefont
  {Rubtsov}},\ }\href {http://stacks.iop.org/0953-8984/27/i=16/a=165601}
  {\bibfield  {journal} {\bibinfo  {journal} {Journal of Physics: Condensed
  Matter}\ }\textbf {\bibinfo {volume} {27}},\ \bibinfo {pages} {165601}
  (\bibinfo {year} {2015})}\BibitemShut {NoStop}%
\bibitem [{\citenamefont {Levy}\ \emph {et~al.}(2007)\citenamefont {Levy},
  \citenamefont {Lahoud}, \citenamefont {Shomroni},\ and\ \citenamefont
  {Steinhauer}}]{Levy2007}%
  \BibitemOpen
  \bibfield  {author} {\bibinfo {author} {\bibfnamefont {S.}~\bibnamefont
  {Levy}}, \bibinfo {author} {\bibfnamefont {E.}~\bibnamefont {Lahoud}},
  \bibinfo {author} {\bibfnamefont {I.}~\bibnamefont {Shomroni}}, \ and\
  \bibinfo {author} {\bibfnamefont {J.}~\bibnamefont {Steinhauer}},\
  }\href@noop {} {\bibfield  {journal} {\bibinfo  {journal} {Nature}\ }\textbf
  {\bibinfo {volume} {449}},\ \bibinfo {pages} {579} (\bibinfo {year}
  {2007})}\BibitemShut {NoStop}%
\bibitem [{\citenamefont {Flory}(1969)}]{Flory1969}%
  \BibitemOpen
  \bibfield  {author} {\bibinfo {author} {\bibfnamefont {P.~J.}\ \bibnamefont
  {Flory}},\ }\href@noop {} {\emph {\bibinfo {title} {Statistical mechanics of
  chain molecules}}}\ (\bibinfo  {publisher} {Wiley, New York},\ \bibinfo
  {year} {1969})\BibitemShut {NoStop}%
\end{thebibliography}%

\end{document}